# Shape-dependent local strain in gold nanorods: data-driven atomic-resolution electron microscopy analysis

Kohei Aso*,1,2, Jens Maebe3, Xuan Q. Tran1, Tomokazu Yamamoto4, Yoshifumi Oshima2, and Syo Matsumura1,4

1. Department of Applied Quantum Physics and Nuclear Engineering, Kyushu University, Nishi-ku, Fukuoka, 819-0395, Japan
2. School of Materials Science, Japan Advanced Institute of Science and Technology, 1-1 Asahidai, Nomi, Ishikawa, 923-1292, Japan
3. Faculty of Sciences, Ghent University, Ghent, B-9000, Belgium
4. The Ultramicroscopy Research Center, Kyushu University, Nishi-ku, Fukuoka, 819-0395, Japan

*Corresponding author's E-mail: aso@jaist.ac.jp

## ABSTRACT

The local variation in inter-atomic distances, or local lattice strain often influences significantly material properties of nanoparticles. Strain measurement with ∼±1% precision is provided by recent atomic-resolution electron microscopy. However, the precision has been limited by noises in the experimental data. Here, we have applied one of the data-driven analyses, Gaussian process regression to predict true form of strain. The precision has been improved to be sub-percent of ±0.2 % and more for detection of local strain. Rod-shaped nanoparticles have been revealed to contain characteristic lattice expansion ∼+0.6 % around the subsurface cap tip area.

The experimental results are reproduced by molecular dynamics simulations of the corresponding shaped atomic models. The strain peculiar to nanorods are explained in terms of curvature-dependent non-uniform surface stress due to shape anisotropy. The present results bring a hint to nanoscale engineering to optimize the strain in nanoparticles by shape control.

**MAIN TEXT**

**Introduction**

Metal nanoparticles attract interests in the wide-areas, such as biology[1], medicine[2], chemistry[3], and physics[4–6] because characteristic properties depend on the size and shape. Nanoparticles with various morphologies are available through chemical synthesis[7], beyond the crystallographically stable shape known as Wolff polyhedron[8]. Especially, gold nanorods are extendedly studied since their chemical stability and near-infrared optical properties depending on the isotropic shape [6]. As well as the morphology, local variation of the atomic distance, or local lattice strain affects the catalytic[9–11] and optical properties[12,13] of metal nanoparticles. The morphology and local strain characterization of metal nanoparticles within high spatial resolution and precision is one of the important keys to understand the mechanisms of the properties and for their improvement.

Transmission electron microscopy (TEM) and scanning TEM (STEM) has been widely utilized to reveal local lattice distortion in nanoparticles below 10 nm in size[14–20], such as surface relaxation [14,15], lattice mismatch at the core-shell bimetallic interface [16,17], lattice expansion due to twinned multi-domain [18,19], strain at the interface of supporting materials[20], and so on. Internal local lattice strain is expected when a nanoparticle gets an anisotropic shape such as rod-like, owing to local change in surface curvature or in surface tension [21]. However, the shape induced internal lattice strain in nanoparticles has not been investigated experimentally, probably because both sub-nanometer spatial resolution and sub-picometer precision are needed.

The strain analysis from atomically resolved (S)TEM images can be divided into mainly two methods; analysis in reciprocal space and real space. Local strain in real space results in intensity broadening of Fourier power components of images in reciprocal space. The local strain can be reconstructed from the Fourier power components by a geometric phase analysis (GPA) [22]. GPA achieves detection of local strain due to local atom displacements of 3 pm [23] to 1 pm [24] in (S)TEM images. In contrast, the real space method detects directly local strains from atomic positions in high-angle annular dark-field (HAADF) imaging of STEM [25]. But the precision obtained in the latter method can be readily influenced by image noises and image distortion due to the instability of the equipment as well as the specimen drift during image acquisition. The image distortion can be suppressed considerably by integrating plural HAADF images acquired in a short time [14,26]. Achieved precisions in atom location are reported to be 5 pm [26] to 0.8 pm [14] for real-space analysis of HAADF images. However, the displacements should be differentiated to evaluate the local strain, which is disturbed sensitively by image noises.

Here, we applied data-driven analysis in atomic resolution HAADF imaging to improve the precision in evaluation of local lattice strain in gold nanorods in real space. It has resulted in such a high accuracies as ~ 1.0 pm in atom location and ~0.2% in lattice strain. A gold nanosphere showed only surface contraction owing to its shape isotropy, while nanorods contained dilatational strain along the long axis in the cap tip parts. The strain magnitudes were quantified to be ~ +0.6% along the axial direction and ~ –0.4% along the width direction of the rods. The local strain is discussed with the support of molecular-dynamics (MD) structural relaxation simulations, and is explained in terms of the curvature-dependent surface stress and shape anisotropy.

## Results and Discussions

**High precision atomic resolution imaging**

Figure 1a-c show atomically resolved HAADF images for three single-crystalline gold nanoparticles that have similar widths of $w$ ~ 9 nm but different lengths of $l$ ~ 9 nm, 21 nm, and 36 nm (The details are available in methods). Here the viewing direction is [$\bar{1}$10] and the $x$- and $y$-directions of the images are set along [001] and [110] of the length and width directions, respectively. The three particles are hereafter referred as NP1.0 (a), NP2.1 (b), and NP3.6 (c), according to the aspect ratios AR = $l/w$ of 1.0, 2.1, and 3.6, respectively. While NP1.0 is in an almost isotropic shape, NP2.1 and NP3.6 are in anisotropic rod-shapes, which consist approximately of cylindrical bodies capped with hemispherical tips at both ends. To improve the signal-to-noise ratio and to minimize image distortions, the images in Fig. 1 were obtained in 2048 × 2048 pixels by integrating the multiple 10–40 frames acquired from the same areas with a fast acquisition speed of 1 μsec/pixel. Distortion of entire images due to specimen drift was corrected by affine transformations that are estimated from the drift rate and acquisition time[27,28]. The irregularities of probe scanning also result in distortion of an entire image and was corrected with images of cubic SrTiO$_3$ [27,28] obtained under the same STEM condition (The details are available in SI#1 and SI#2).

Atomic column positions were determined precisely as the peak coordinates of 2-dimensional Gaussian functions[29] fitted to the image intensity profiles for bright spots in the HAADF images [30]. The distances between neighboring atom columns were then measured along [110] and [001] directions. The intercolumnar distances $d_{110}$ and $d_{001}$ are plotted in Figures 1d and 1e as a function of the aspect ratios of particles, respectively. Here, the particle interiors were divided into two areas: the core and the shell regions. The former is defined as the central rectangle area with a width of $w/2$ and a length of $l/2$, while the latter is the rest part including the surface and the tip areas. The averaged distances of $d_{110}$ and $d_{001}$ agree within the standard deviations (error bars) with the values indicated by horizontal gray lines obtained from bulk lattice parameter $a_{bulk}$ = 407.84 pm [31], both in the core and in the shell regions. It proves that the image distortions have been successfully corrected. The standard deviations for the core regions extend only in the range of 2.3–4.6 pm, which is smaller than the image pixel sizes of 8.3–27.2 pm/pixel. It indicates that sub-pixel and picometer-order precision has been achieved in location of atomic columns in the images. The standard deviations are expanded slightly by ~1 to 2 pm for the shell regions from the cases for the cores, suggesting existence of local lattice distortion in the shell regions .

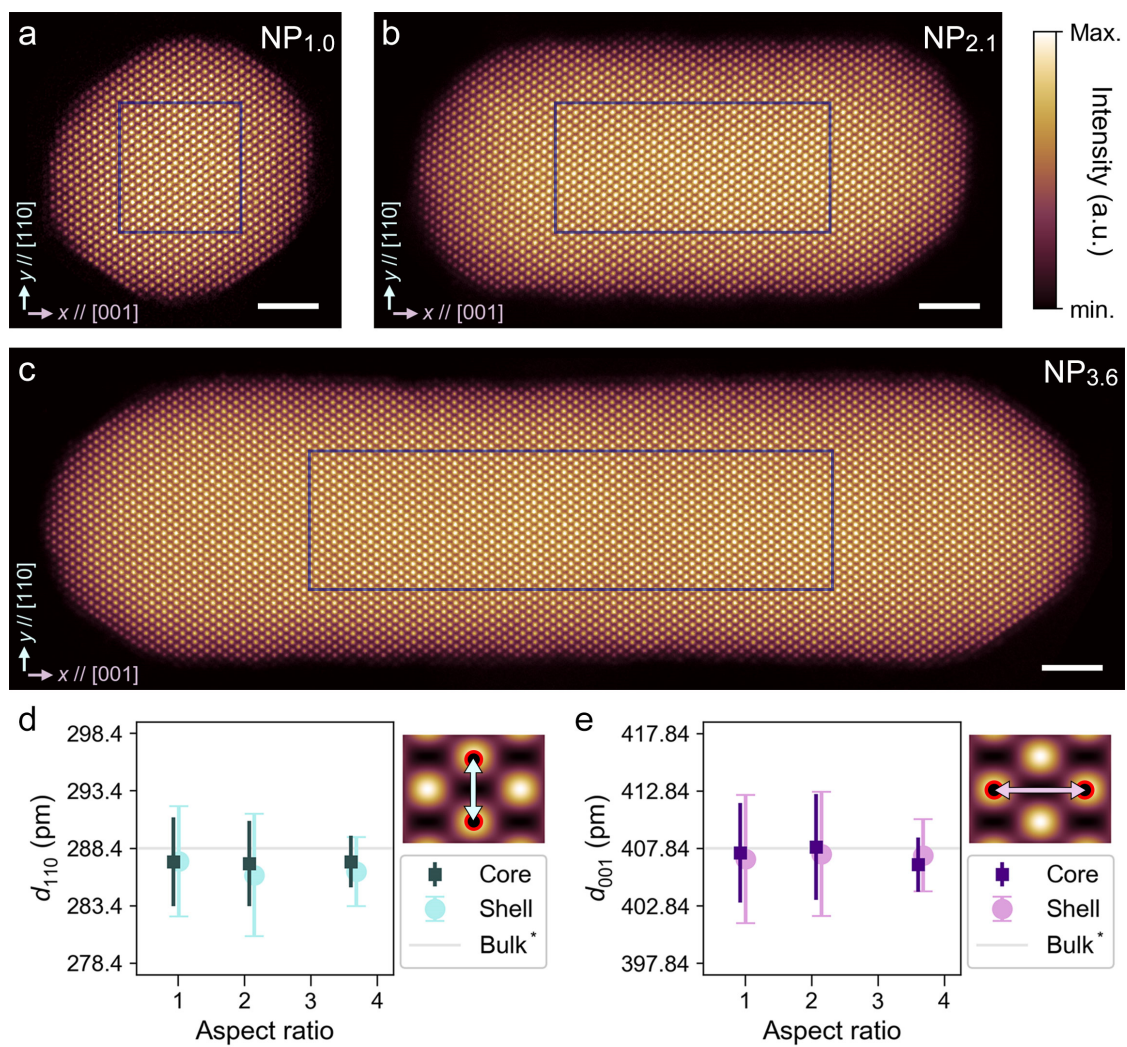

**Figure 1. Atomically resolved experimental images and the statistics of atomic distances.** HAADF images of the particles with similar widths ($w \sim 9.0$ nm) but different AR $\sim 1.0$, 2.1, and 3.6 ($l$: length), namely **(a)** $NP_{1.0}$, **(b)** $NP_{2.1}$, and **(c)** $NP_{3.6}$. The [001] and [110] crystal orientations are aligned parallel to the $x$- and $y$-directions of the viewing images. The scale bars correspond to 2 nm. The color scale is in arbitrary units (*a.u.*). The rectangle indicates the core region with length $l/2$ and width $w/2$. The shell region is defined as the other area of the core region. Plots of atomic distances **(d)** $d_{110}$ and **(e)** $d_{001}$ as a function of the aspect ratio of the particles. The squares and circles are the average atomic distances in the core and shell regions. The vertical bars are standard deviations. The horizontal gray lines are gold bulk values from ref [31].

**Data-driven strain analysis**

The displacements of atomic columns were evaluated as deviations of the positions from the reference lattice sites which were obtained by extension of the averaged periodicities in the core regions to the whole areas (See the details in method)[32]. The experimentally obtained values of displacements inevitably include random fluctuations or noise components, which interfere the estimation of local strains by differentiation of local displacements. The Gaussian process regression (GPR) [33,34] was used to remove the random noises in the raw data of displacements $\boldsymbol{u}^{\text{raw}}(x_m, y_n)$ and to obtain the smooth and continuous displacement field $\hat{\boldsymbol{u}}^{\text{GPR}}(x, y)$. GPR is one of the data-driven methods and has been used widely in machine learning processes [33]. GPR takes advantage of Bayesian statistics to reduce the noise components and derives a smooth continuous function fitted to spatially discrete data. The details of GPR are given in SI#3.

Figure 2 illustrates $x$-direction components of $u_x^{\text{raw}}(x_m, y_n)$ and $\hat{u}_x^{\text{GPR}}(x, y)$ for NP2.1. The fine speckled variations in the color scales are found in (a), corresponding to random fluctuations in $u_x^{\text{raw}}(x_m, y_n)$. In contrast, the speckled variations have been removed and the continuous color gradation appears in (b) for $\hat{u}_x^{\text{GPR}}(x, y)$ after GPR. The dark blue and bright yellow color scales of the maps (a) and (b) indicate displacements toward $-x$ and $+x$ directions from the reference periodic lattice, respectively. The outward displacements along $x$ direction of the longer axis can be recognized in the hemispherical caps at the left side and right side tips, respectively. One-dimensional plots of $u_x^{\text{raw}}(x_m, 0)$ and $\hat{u}_x^{\text{GPR}}(x, 0)$ along the central axis are compared to each other in Figure 2c. One may again confirm that GPR has successfully removed random noise components and has disclosed proper atomic displacements with the smooth continuous line for $\hat{u}_x^{\text{GPR}}(x, 0)$. It is noticed again that $\hat{u}_x^{\text{GPR}}(x, 0)$ takes negative and positive values in the left and the right end portions, respectively, about 5 nm and more away from the center. Outward displacements occur both in the end portions, but subside in the tip ends probably by surface contraction. In Figures S3 and S4, one can see that GPR has successfully remove the random noise components in $u_x^{\text{raw}}(x_m, y_n)$ and $u_y^{\text{raw}}(x_m, y_n)$ for the three samples. The outward displacements along *x*-direction are again recognized in both tip portions in rod-shaped NP3.6, but not in NP1.0, as shown in Fig. S3, being regarded as a characteristic feature due to rod-shape. Only small displacements take place limited in a very thin surface layer in NP1.0. Figure S4 shows that displacement fields are set up also in $u_y^{\text{raw}}(x_m, y_n)$ and $\hat{u}_y^{\text{GPR}}(x, y)$. The displacements along *y*-direction occur inward broadly in the tip portions in NP2.1 and NP3.6, but limited only in thin surface skins of NP1.0 as well as of the main cylindrical bodies in NP2.1 and NP3.6. The precision in atomic column locations would be effectively improved to be 1.0–1.6 pm from 2.1–3.1 pm if $\hat{u}_x^{\text{GPR}}(x, y)$ and $\hat{u}_y^{\text{GPR}}(x, y)$ in the core regions are used for the evaluation in place of $u_x^{\text{raw}}(x_m, y_n)$ and $u_y^{\text{raw}}(x_m, y_n)$, because the statistical noises have been minimized via GPR. The displacement fields of $\hat{u}_x^{\text{GPR}}(x, y)$ and $\hat{u}_y^{\text{GPR}}(x, y)$ look enough smooth to evaluate local lattice strains by their partial differentiation in the following sections.

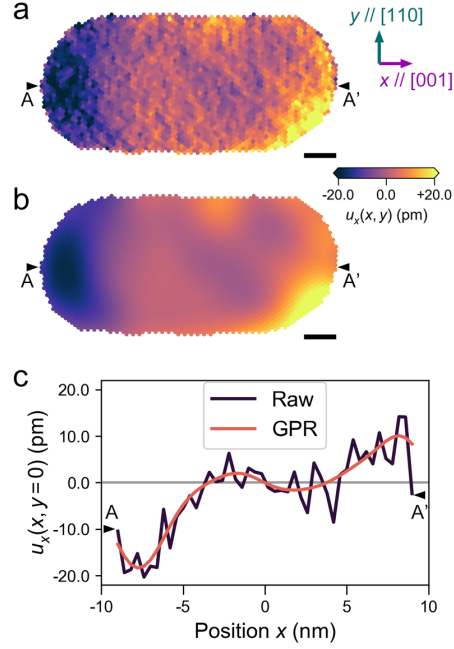

**Figure 2. GPR of experimental displacements.** Raw displacements of atomic positions from the periodic lattice **(a)**, and atomic displacement field obtained by GPR **(b)**. The dark blue and bright yellow colors in the maps correspond to atomic displacements toward –*x* and +*x*. The scale bars correspond to 2 nm. **(c)** gives linear plots of atomic displacements $u_x^{\text{raw}}(x_m, 0)$ and $\hat{u}_x^{\text{GPR}}(x, 0)$ along the central axis.

**Strain along width (*y*)-direction**

Figure 3a shows lattice strain $e_{yy}(x,y)$ maps in the three nanoparticles, which have been obtained by $\partial \hat{u}_y^{\text{GPR}}(x, y)/\partial y$. Although inward displacements along *y*-direction are pronounced broadly in the hemispherical tip regions in the rod shaped particles NP2.1 and NP3.6 as shown in Figure S4, negative $e_{yy}(x,y)$ colored with dark blue appears only the surface skin layers in the main cylindrical portions in Fig. 3a. This suggests that inward lattice shifts along [110] direction are major components in the displacements in the hemispherical tip interior, bearing little lattice strain. This characteristic feature is confirmed in the corresponding molecular dynamics (MD) simulations given in Figure 3b and Figure S4. The details of MD simulations are described in SI #6. After structural relaxation at 300 K, the atomic positions in a model particle were averaged along to [$\bar{1}10$] to imitate the experimental HAADF imaging. Particle models named MDrod are a faceted sphere and rods resembled to the real nanoparticles in terms of size and shape, as shown in Figure S8a-c. One can see that color change for displacements occurs broadly in both side tips and limitedly in thin surface layers in the main cylindrical parts in MDrod for NP2.1 and NP3.6 in Figure S4, while distinct dark blue bands for negative $e_{yy}$ appear only in the top and bottom sub-surfaces in Figure 3. Faint islands of negative $e_{yy}$ are slightly recognized in the interior center of hemispherical tips in b2 and b3 in Figure 3, probably corresponding to a pincer operation of downward and upward lattice slides. Interestingly, the maps of MDsph for ellipsoids in Figure 3 and Figure S4 show inward strain and displacements only in the surface skins due to the smoothly curved surface. The local surface curvature change between cylindrical and spherical modes in rod-shaped particles is responsible for the displacements and strain caused in the cap interiors. It should be noted that a rough stripe pattern faintly appears in the particle interior of NP3.6, as seen in Figure 3(a3). The stripes are most likely to artifacts due to nonsquareness in the electron probe scanning still residual even after distortion correction, because they appear parallel to the scan direction of electron probe. The nonsquareness in the electron probe scanning and its influences to local strain evaluation are discussed with a cubic lattice $SrTiO_3$ specimen in SI #5. The striped modulations in Figure 3a3 amounts to ±0.13 % of strain $e_{yy}$, which can be regarded as the detection limit or the uncertainty in strain evaluation in the present study.

Figure 3 (d1-d3) compare local variations of $\bar{e}_{yy}(y)$ averaged over *x*-direction in the experimental, MDrod and MDsph maps. Here, $\bar{e}_{yy}(y)$ was obtained by

$$\bar{e}_{yy}(y) = \frac{\int_{-l/2}^{l/2} e_{yy}(x,y)dx}{\int_{-l/2}^{l/2} A(x,y)dx} \quad (1)$$

with an area function $A(x,y)$ which takes unity inside the particle area but otherwise zero. The corresponding averaged variations in $\bar{u}_y(y)$ are given in Figure S4 (e1-e3). The dotted and dashed lines for MDrod and MDsph show distinctly contract strain occurring in the surface skins about 1 nm thick at the left and right sides, and weak oscillations in $\bar{e}_{yy}(y)$ in the interiors. The surface contraction becomes weaker in MDrod than in MDsph, probably owing to partially flattened habit surface in the former model, as seen in Figure S8. The solid lines for the experimental are qualitatively in agreement with the simulated lines of MDrod and MDsph models. Only weaker surface contraction is obtained in the experimental results than in the corresponding simulations. But we are enough satisfied with the accuracies achieved in the experimental because local roughness in the real surfaces may disturb and shade off the uniform surface strain. The weak dilatation fields can be recognized just inside the contracting surface skins in the experimental solid lines in d1 and d2, as predicted by the simulations.

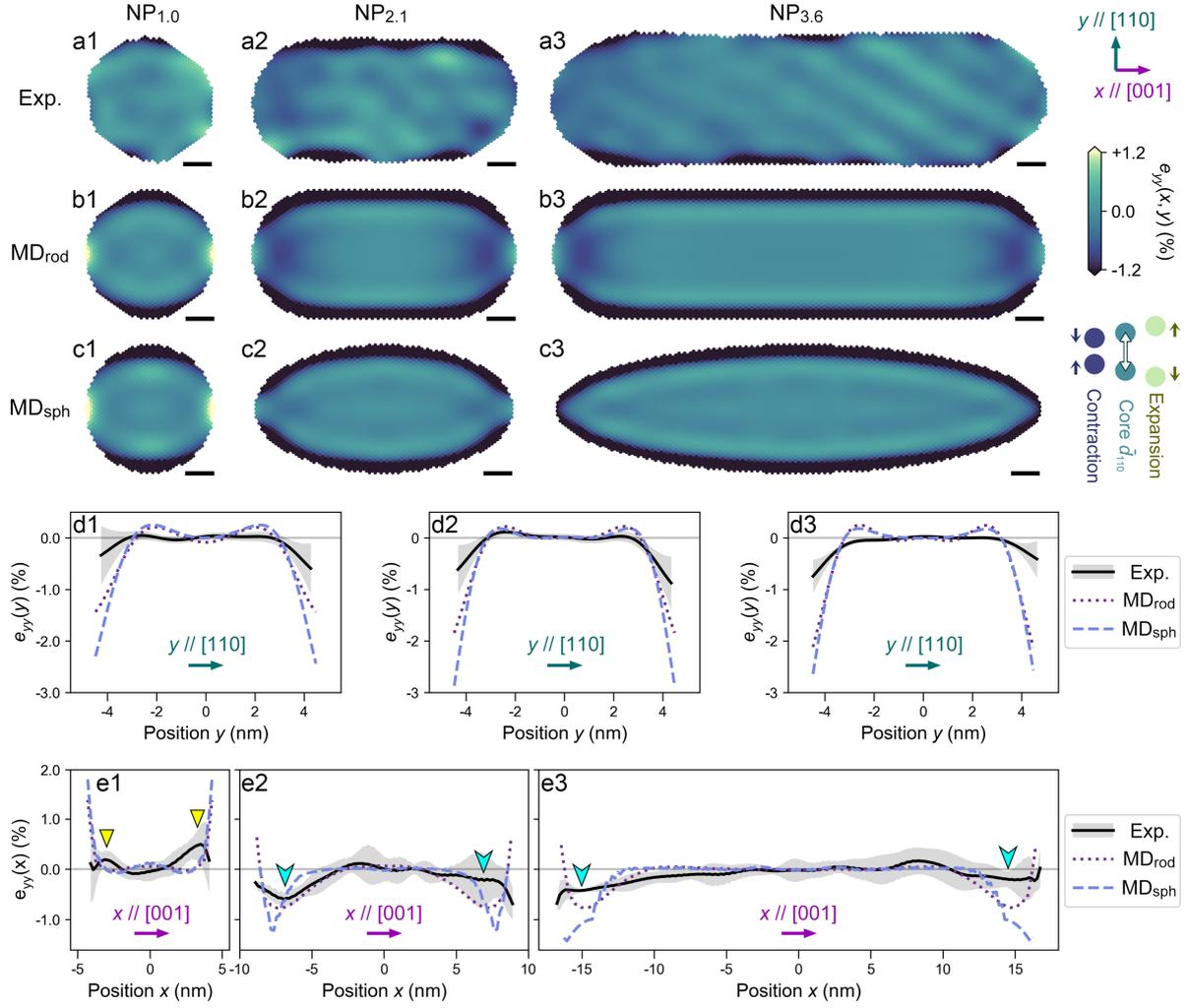

**Figure 3. Local distribution of lattice strain along width (*y*-) directions of particles.** Two-dimensional maps of strain $e_{yy}(x,y)$ for **(a)** experiments, simulated **(b)** MD$_{rod}$, and **(c)** MD$_{sph}$. The color corresponds to strain along *y*-direction as schematically illustrated. Each scale bar corresponds to 2 nm. Line profiles of strain **(d)** $e_{yy}(y)$ and **(e)** $e_{yy}(x)$, where the strain is averaged over each {110} central layer with a width of *w*/2 and {001} central layer with a length of *l*/2. The experimental curves are shown together with their statistical 95% confidential intervals. Index 1, 2, and 3 following the captions mean particles with the aspect ratios of 1.0, 2.1, and 3.6.

**Strain along length (*x*)-direction**

Figure 4 shows maps of lattice strain $e_{xx}(x,y)$ directed to the long axis in the three nanoparticles evaluated by $\partial \hat{u}_x^{GPR}(x,y)/\partial x$, in comparison with the simulations for MDrod and MDsph models. The original displacement fields of $u_x(x,y)$ are illustrated in Figure S4.

First of all, one may notice that positive $e_{xx}(x,y)$ regions colored with light yellow appear in the tip end portions of both right and left sides in the experimental and MDrod maps for NP2.1 and NP3.6, in contrast to no light yellow fields in isotropic spheres NP1.0 as well as in ellipsoids of MDsph models. The local displacements $u_x(x,y)$ in the tip parts are set up in completely opposite sense between rod-shaped particles and ellipsoids of MDsph, as seen in Figure S4. It is clearly revealed from Figures 3 and 4 that spherical and spheroidal nanoparticles involve only contracting strain in the sub-surface layers, while rod-shaped nanoparticles include dilatation strain along to the long axis (*x*) direction localized in hemispherical cap interior in addition to the contracting surface skins.

The characteristic strain fields are confirmed quantitatively in strain lines of averaged $\bar{e}_{xx}(x)$ in Figure 4 d1-d3 as well as lines of $\bar{u}_x(x)$ in Figure S4 e1-e3. Here $\bar{e}_{xx}(x)$ and $\bar{u}_x(x)$ were obtained through *y*-integration in an analogous way to equation (1). The three lines for NP1.0 and the dashed lines for ellipsoids of NP2.1 and NP3.6 are drawn within the negative $\bar{e}_{xx}(x)$ range. In contrast, the solid and dotted lines get increased into the positive side about 5 nm inside from the both ends and are then dropped down into negative at the tip ends after drawing a peak about 0.5 %, as shown in Figure 4 d2 and d3. In the simulation maps for NP2.1 and NP3.6, the cylindrical main parts of particle interior are quite uniform as null, while mottled irregular patterns faintly appear in the corresponding areas in the experimental maps. The irregular variations are in a range between ± 0.16 %, which can be regarded as uncertainty in strain determination in the present approach.

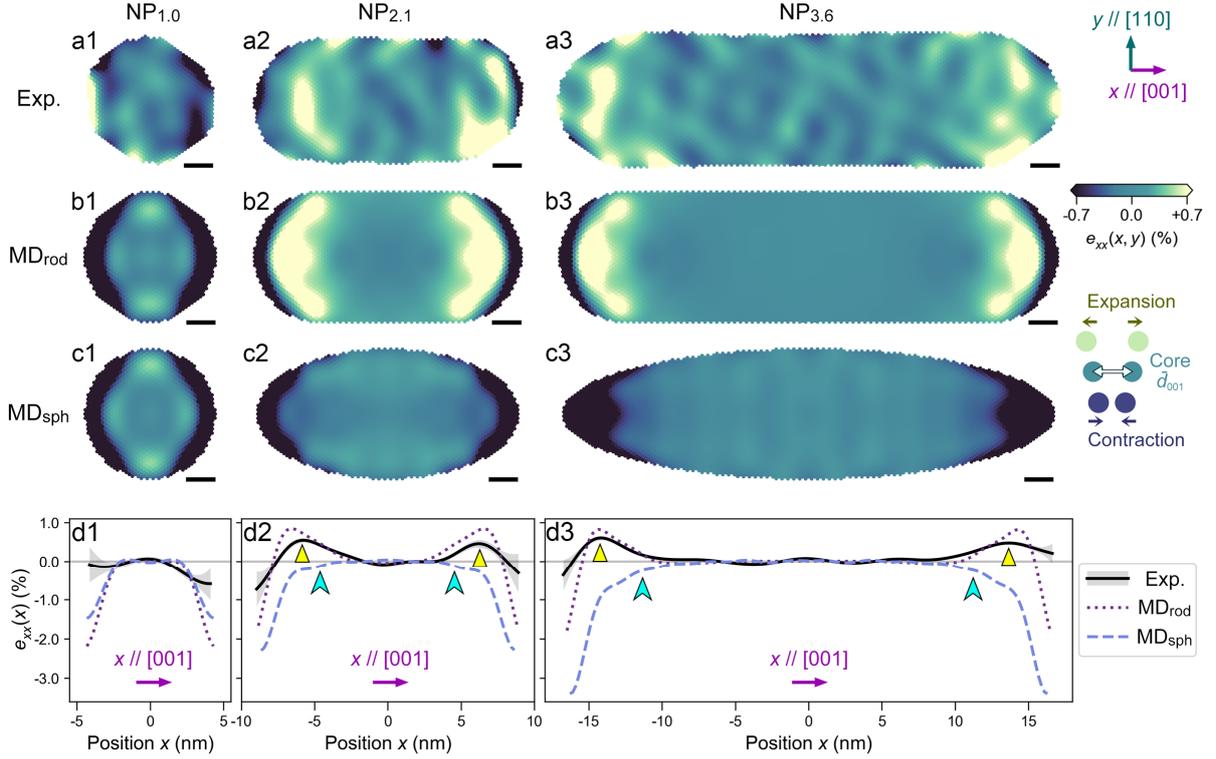

**Figure 4. Local distribution of lattice strain along with length directions of particles.** Two-dimensional maps of strain $e_{xx}(x,y)$ for **(a)** experiments, simulated **(b)** $MD_{rod}$, and **(c)** $MD_{sph}$. The color corresponds to strain along *y*-direction as schematically indicated in center-right. Each scale bar corresponds to a length of 2 nm. Line profiles of strain **(d)** $e_{xx}(x)$, where the strain is averaged over each {001} layer, as schematically indicated in the panels. The experimental curves are shown together with their statistical 95% confidential intervals. Index 1, 2, and 3 following the caption mean particles with the aspect ratios of 1.0, 2.1, and 3.6.

**The origin of strain in nanorods**

The strain distribution in nanorods indicates the non-uniform surface stress. The stress is related to the curvature of a surface by Young-Laplace (YL) equation [35,36] that is also used for solid-vacuum interface[37,38]. The main concept of the YL equation is that the amount of normal stress $\sigma_n$ (N) relates to the curvature $H$ (m$^{-1}$) of a surface [39]:

$$\sigma_n = 2sH \qquad (1),$$

where $s$ = 1.175 (N/m) is the surface stress of gold [40]. The value is obtained by Mays *et al* [38] through YL equation: they also reported the size-dependent volume contraction of spherical gold nanoparticles, resulting from the inward stress as schematically illustrated in Figure 5a.

The curvature-dependent stress is the main cause of the local strain in nanorods. Here, we assumed a rod model consisted of a cylindrical body and hemispherical cap ends, with mean curvatures of $H_{bod}$ = 1/w on the body and $H_{cap}$ = 2/w on the caps. In the illustrated stress of a nanorod in Figure 5b-c, the normal stress $\sigma_n$ (Figure 5b) is decomposed into the components $\sigma_r$ along the radial direction (Figure 5 エラー! 参照元が見つかりません。c) and $\sigma_x$ along the axial direction (Figure 5d). Here, the cylindrical coordinate system is used in consideration of the rotational symmetry around the long axis. The stress component $\sigma_r$ increases twice at the cap-side boundary, compared to the body region, as seen in Figure 5c.

Shape anisotropy is the reason why the strain $e_{rr}$ is significant compared to the strain $e_{xx}$. Focusing on the cap region, the surface stress acts similarly to the sphere: the stress $\sigma_r$ is relaxed only for hemisphere volume. The stress $\sigma_x$ is also applied similarly in Figure 5d, however, the volume to which stress is applied is different: the stress $\sigma_x$ is relaxed not only for hemisphere volume but also for the whole body volume. Therefore, the lattice around the boundary contracts along the radial (*r*-) direction, allowing the local lattice expansion along the axial (*x*-) direction through the Poisson strain. The ratio of experimental strain at the cap ends is calculated to be $|e_{yy, cap}|$ / $|e_{xx, cap}|$ = 0.7, where $|e_{xx, cap}|$ ~ 0.5% is the expansion along *x*-direction, and $|e_{yy, cap}|$ ~ 0.35% is the contraction along *y*-direction. The ratio corresponds to the Poisson's ratio of [001] for longitudinal and [110] for lateral directions, $v$(001, 110) = 0.46 for bulk gold [32,43].

Strain due to curvature-dependent stress is evaluated to be enough comparable through the above model and Hooke's law. From mean curvature of a sphere $H_{cap} = 2/w \sim 0.22$ nm$^{-1}$ ($w = 9.0$ nm) and Equation (1), the amount of stress around the boundary is evaluated to be $\sigma_r = 2sH_{cap} = 520$ MPa. The stress $\sigma_r$ can convert into strain $e_{rr}$ from the Hooke's law

$$e_{rr} = \frac{\sigma_r}{E} \tag{2},$$

where $E = 79$ GPa is Young's modulus of the bulk gold [31]. The obtained amount of contraction strain $|e_{rr}| \sim 0.66\%$ corresponds to the amount $|e_y| \sim -0.2$ to $-0.5\%$ observed around the cap region in the experimental rods (Figure 3e2 and e3). The strain amount is determined by the curvature at the hemisphere, which also corresponds to the experimental trends that the strain amount is not highly dependent on the aspect ratio of nanorods (Figure 4d2 and d3).

Curvature dependent stress is also applied to explain the contraction strain in spheroids of MD2 simulations. The curvature of a spheroid changes continuously on its surface

$$H(x) = \frac{l\{w^2 + l^2 + (AR^2 - 1)4x^2\}}{\{l^2 + (AR^2 - 1)4x^2\}} \tag{3},$$

where $w$ is the width, $l$ is the length along the x-direction, AR = $l/w$ is the aspect ratio, and the center of the spheroid locates at $x = 0$. Resulting stress distribution around a spheroid is schematically illustrated in Figure 5e. A spheroid takes curvature maximum of

$$H_{Max} = \frac{2l}{w^2} = \frac{2AR}{w} \tag{4}$$

at the ends of a spheroid ($x = \pm l/2$). As the $AR$ increases, the curvature maximum $H_{Max}$ increases, resulting in more stress at the ends of a spheroid along the long axis. This corresponds to the simulation results; spheroids exhibit the huge local contraction at the end surfaces, and the contraction amount seems dependent on the aspect ratio (Figure 4d2 and d3).

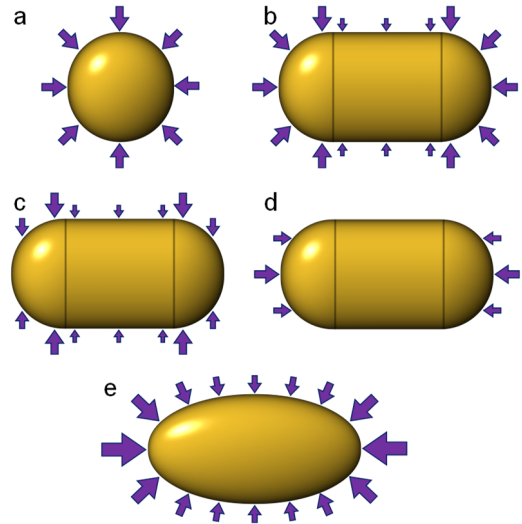

**Figure 5. Curvature dependent stress in anisotropic shaped nanoparticles.** A sphere **(a)**, a rod **(b)**, $\sigma_r$ components **(c)**, and $\sigma_x$ component on a rod **(d)**, a spheroid **(e)**.

## Conclusions

We have improved location of atomic column positions by atomic-resolution scanning transmission electron microscopy to be about 1.0 pm in precision by taking advantage of data-driven GPR processing of statistical noise components as well as correction on probe-scanning distortion. It enables us to detect local lattice strain in magnitude of 0.2% and larger in gold nanoparticles. In the present study, this technique has been applied to characterize the shape effect on internal local strain in gold nanoparticles and has revealed that rod-shaped nanoparticles contain characteristic dilatation strain about +0.6 % along the rod axis direction localized in the end cap interior. The experimentally obtained results have been reproduced by the corresponding molecular dynamics (MD) simulations. The strain peculiar to nanorods is explained in terms of curvature-dependent non-uniform surface stress due to shape anisotropy. The present results demonstrate well direct evidence and understanding of shape effect on local interior strain induced in nanoparticles, and give a hint to nanoscale engineering to optimize the strain in nanoparticles by shape control.

## Methods

**Sample preparation:** Gold nanorods used in this study were obtained from a commercial lot of averaged length and diameter of 50 and 10 nm (products of Dai Nihon Toryo Co. Ltd., Japan), which were produced in a cetyltrimethylammonium bromide (CTAB) micellar solution by a photochemical method. Their optical absorption spectra showed peaks at wavelengths of ~520 and 980 nm. The CTAB micelles were partially removed from the solution by centrifuging for 10 min at $2 \times 10^4 g$ (where $g$ is the gravitational acceleration). One drop of the centrifuged solution was added onto a Quantifoil™ carbon film supported on a grid with a thickness of approximately 20 nm. The grid had been rendered hydrophilic after exposure to argon ions in a plasma cleaner for 3 min.

**Image acquisition under drift distortion correction:** HAADF observations of gold nanoparticles oriented to $[\bar{1}10]$ direction were carried out using a STEM with a cold field-emission gun (JEM-ARM200F ACCELARM, JEOL, Japan). The acceleration voltage of an electron beam was set to be 120 kV lowered down from the nominal voltage 200 kV, to reduce atom displacements due to electron illumination. The convergence semi-angle of the incident electron probe was 19 mrad, which corresponds to the depth resolution of ~ 19 nm [41]. The angular detection range of the HAADF detector for the scattered electrons was 70–150 mrad. The observations were conducted with fast scan speed by a customized script for Digital Micrograph™ to suppress the influence of sample drift during STEM operation. A series of rapidly scanned HAADF images were acquired for an area of interest with a dwell time as short as 1 μs/pixel. The image resolutions were adjusted accordingly to the particle size: 512 × 512 pixels for NP1, and 2048 × 2048 pixels for both NP2 and NP3 samples. Image distortions are corrected via the affine transformation[27,28]. To correct image distortion due to specimen drift, the affine matrix for transformation was deterimed on image acquisition conditions such as drift rate, scanning fly-back time, and acquisition interval. Systematic error due to STEM equipment was corrected from the calibration of a standard sample SrTiO$_3$ (See SI#2). By integrating the corrected image series of 10–40 frames, the final images of nanoparticles were then constructed.

**Calculation of atomic displacements:** Gaussian functions were fitted to the image intensity of atomic columns that appeared as peaks in a HAADF image [30]. The function fitting has the advantage of being less susceptible to variations in image intensity between adjacent pixels due to noise and enables to determine peak positions within subpixel scale. The fitting procedure was carried out with a software StatSTEM [29]. The peak-top positions of the fitted Gaussians were defined to be atomic positions. Each atomic column position, characterized by a coordinate vector $\boldsymbol{r}_\text{exp} = (x_m, y_n)$, was determined by fitting a two-dimensional Gaussian function to the intensity profile of the experimental HAADF images[29]. The periodic reference lattice $\boldsymbol{r}_\text{ref} = (x'_m, y'_n) = \boldsymbol{O} + \frac{m}{2}\boldsymbol{p}_{[001]} + \frac{n}{2}\boldsymbol{q}_{[110]}$ was prepared with integers $m$ and $n$ for the indexes to express the $(\bar{1}10)$ periodic lattice. The parameter $\boldsymbol{O}$ is the origin, and $\boldsymbol{p}_{[001]}$ and $\boldsymbol{q}_{[110]}$ are the lattice vectors in the [001] and [110] directions, respectively. The reference lattice and corresponding experimental atom positions $\boldsymbol{r}_\text{exp}$ were superimposed, and the parameters $\boldsymbol{O}$, $\boldsymbol{p}_{[110]}$ and $\boldsymbol{q}_{[001]}$ were determined by minimizing the residual sum of square (RSS) of $\sum_m \sum_n \|\boldsymbol{r}_\text{exp} - \boldsymbol{r}_\text{ref}\|^2$ in the core region that is defined in the main text. The optimization process was carried out by using Basin-hopping algorithm to reach a global

minimum of the RSS[42]. The original data of atom displacements were collected as $\boldsymbol{u}_\text{raw}(x_m, y_n) = \boldsymbol{r}_\text{exp} - \boldsymbol{r}_\text{ref}$ using the optimized parameters[32].

## ASSOCIATED CONTENT

### Supporting Information

Supporting Information is available free of charge on the ACS Publications website. See supporting information for more detail on the sample preparation, HAADF observation, strain analysis, evaluation of error, detailed conditions of the MD simulations.

## AUTHOR INFORMATION

### Corresponding Author

*E-mail: aso@jaist.ac.jp

### Orchid

Kohei Aso: 0000-0001-6935-7655

### Competing Financial Interests

The authors declare no competing financial interests.

## ACKNOWLEDGEMENTS

The authors are indebted to Prof. Yukio Sato at the Kyushu University for a discussion of image distortion correction, and Dr. Masaki Kudo and Mr. Takaaki Toriyama at the Ultramicroscopy Research Center for technical supports. This work was supported partly by Grant-in-Aid for Scientific Research B (No. 25289221 and No. 18H01830) from Japan Society for the Promotion of Science, the ACCEL program (JPMJAC1501), Japan Science and Technology Agency (JST)and Education and Research Support Program on Mathematics and Data Science 2017 from Kyushu University.

# REFERENCES

(1) Yang, L.; Zhou, Z.; Song, J.; Chen, X. Anisotropic Nanomaterials for Shape-Dependent Physicochemical and Biomedical Applications. *Chem. Soc. Rev.* **2019**, *48* (19), 5140–5176. https://doi.org/10.1039/c9cs00011a.

(2) Kinnear, C.; Moore, T. L.; Rodriguez-Lorenzo, L.; Rothen-Rutishauser, B.; Petri-Fink, A. Form Follows Function: Nanoparticle Shape and Its Implications for Nanomedicine. *Chem. Rev.* **2017**, *117* (17), 11476–11521. https://doi.org/10.1021/acs.chemrev.7b00194.

(3) Ishida, T.; Murayama, T.; Taketoshi, A.; Haruta, M. Importance of Size and Contact Structure of Gold Nanoparticles for the Genesis of Unique Catalytic Processes. *Chem. Rev.* **2020**, *120* (2), 464–525. https://doi.org/10.1021/acs.chemrev.9b00551.

(4) Alloyeau, D.; Ricolleau, C.; Mottet, C.; Oikawa, T.; Langlois, C.; Le Bouar, Y.; Braidy, N.; Loiseau, A. Size and Shape Effects on the Order-Disorder Phase Transition in CoPt Nanoparticles. *Nat. Mater.* **2009**, *8* (12), 940–946. https://doi.org/10.1038/nmat2574.

(5) Molleman, B.; Hiemstra, T. Size and Shape Dependency of the Surface Energy of Metallic Nanoparticles: Unifying the Atomic and Thermodynamic Approaches. *Phys. Chem. Chem. Phys.* **2018**, *20* (31), 20575–20587. https://doi.org/10.1039/c8cp02346h.

(6) Pérez-Juste, J.; Pastoriza-Santos, I.; Liz-Marzán, L. M.; Mulvaney, P. Gold Nanorods: Synthesis, Characterization and Applications. *Coord. Chem. Rev.* **2005**, *249* (17-18 SPEC. ISS.), 1870–1901. https://doi.org/10.1016/j.ccr.2005.01.030.

(7) Grzelczak, M.; Pérez-Juste, J.; Mulvaney, P.; Liz-Marzán, L. M. Shape Control in Gold Nanoparticle Synthesis. *Chem. Soc. Rev.* **2008**, *37* (9), 1783–1791. https://doi.org/10.1039/b711490g.

(8) Wulff, G. Zur Frage Der Geschwindigkeit Des Wachsthums Und Der Auflösung Der Krystallflächen. *Zeitschrift für Krist. - Cryst. Mater.* **1901**, *34* (1–6), 449–530. https://doi.org/10.1524/zkri.1901.34.1.449.

(9) Stamenkovic, V.; Mun, B. S.; Mayrhofer, K. J. J.; Ross, P. N.; Markovic, N. M.; Rossmeisl, J.; Greeley, J.; Nørskov, J. K. Changing the Activity of Electrocatalysts for Oxygen Reduction by Tuning the Surface Electronic Structure. *Angew. Chemie Int. Ed.* **2006**, *45* (18), 2897–2901. https://doi.org/10.1002/anie.200504386.

(10) Moseley, P.; Curtin, W. A. Computational Design of Strain in Core-Shell Nanoparticles for Optimizing Catalytic Activity. *Nano Lett.* **2015**, *15* (6), 4089–4095.

https://doi.org/10.1021/acs.nanolett.5b01154.

(11) Strasser, P.; Koh, S.; Anniyev, T.; Greeley, J.; More, K.; Yu, C.; Liu, Z.; Kaya, S.; Nordlund, D.; Ogasawara, H.; Toney, M. F.; Nilsson, A. Lattice-Strain Control of the Activity in Dealloyed Core–Shell Fuel Cell Catalysts. *Nat. Chem.* **2010**, *2* (6), 454–460. https://doi.org/10.1038/nchem.623.

(12) Park, H. S.; Qian, X. Surface-Stress-Driven Lattice Contraction Effects on the Extinction Spectra of Ultrasmall Silver Nanowires. *J. Phys. Chem. C* **2010**, *114* (19), 8741–8748. https://doi.org/10.1021/jp100456p.

(13) Ben, X.; Park, H. S. Strain Engineering Enhancement of Surface Plasmon Polariton Propagation Lengths for Gold Nanowires. *Appl. Phys. Lett.* **2013**, *102* (4). https://doi.org/10.1063/1.4790293.

(14) Yankovich, A. B.; Berkels, B.; Dahmen, W.; Binev, P.; Sanchez, S. I.; Bradley, S. A.; Li, A.; Szlufarska, I.; Voyles, P. M. Picometre-Precision Analysis of Scanning Transmission Electron Microscopy Images of Platinum Nanocatalysts. *Nat. Commun.* **2014**, *5* (May), 1–7. https://doi.org/10.1038/ncomms5155.

(15) Huang, W. J.; Sun, R.; Tao, J.; Menard, L. D.; Nuzzo, R. G.; Zuo, J. M. Coordination-Dependent Surface Atomic Contraction in Nanocrystals Revealed by Coherent Diffraction. *Nat. Mater.* **2008**, *7* (4), 308–313. https://doi.org/10.1038/nmat2132.

(16) Mukherjee, D.; Gamler, J. T. L.; Skrabalak, S. E.; Unocic, R. R. Lattice Strain Measurement of Core@Shell Electrocatalysts with 4D-STEM Nanobeam Electron Diffraction. *ACS Catal.* **2020**. https://doi.org/10.1021/acscatal.0c00224.

(17) Gamler, J. T. L.; Leonardi, A.; Sang, X.; Koczkur, K. M.; Unocic, R. R.; Engel, M.; Skrabalak, S. E. Effect of Lattice Mismatch and Shell Thickness on Strain in Core@shell Nanocrystals. *Nanoscale Adv.* **2020**, *2* (3), 1105–1114. https://doi.org/10.1039/d0na00061b.

(18) Johnson, C. L.; Snoeck, E.; Ezcurdia, M.; Rodríguez-González, B.; Pastoriza-Santos, I.; Liz-Marzán, L. M.; Htch, M. J. Effects of Elastic Anisotropy on Strain Distributions in Decahedral Gold Nanoparticles. *Nat. Mater.* **2008**, *7* (2), 120–124. https://doi.org/10.1038/nmat2083.

(19) Goris, B.; De Beenhouwer, J.; De Backer, A.; Zanaga, D.; Batenburg, K. J.; Sánchez-Iglesias, A.; Liz-Marzán, L. M.; Van Aert, S.; Bals, S.; Sijbers, J.; Van Tendeloo, G. Measuring Lattice Strain in Three Dimensions through Electron Microscopy. *Nano Lett.* **2015**, *15* (10), 6996–7001. https://doi.org/10.1021/acs.nanolett.5b03008.

(20) Nilsson Pingel, T.; Jørgensen, M.; Yankovich, A. B.; Grönbeck, H.; Olsson, E. Influence of Atomic Site-Specific Strain on Catalytic Activity of Supported
22

Nanoparticles. *Nat. Commun.* **2018**, *9* (1), 1–14. https://doi.org/10.1038/s41467-018-05055-1.

(21) Shuttleworth, R. The Surface Tension of Solids. *Proc. Phys. Soc. Sect. A* **1950**, *63* (5), 444–457. https://doi.org/10.1088/0370-1298/63/5/302.

(22) Hÿtch, M. J.; Snoeck, E.; Kilaas, R. Quantitative Measurement of Displacement and Strain Fields from HREM Micrographs. *Ultramicroscopy* **1998**, *74* (3), 131–146. https://doi.org/10.1016/S0304-3991(98)00035-7.

(23) Hÿtch, M. J.; Putaux, J.-L.; Pénisson, J.-M. Measurement of the Displacement Field of Dislocations to 0.03 Å by Electron Microscopy. *Nature* **2003**, *423* (6937), 270–273. https://doi.org/10.1038/nature01638.

(24) Zhu, Y.; Ophus, C.; Ciston, J.; Wang, H. Interface Lattice Displacement Measurement to 1 Pm by Geometric Phase Analysis on Aberration-Corrected HAADF STEM Images. *Acta Mater.* **2013**, *61* (15), 5646–5663. https://doi.org/10.1016/j.actamat.2013.06.006.

(25) Lee, S.; Oshima, Y.; Hosono, E.; Zhou, H.; Takayanagi, K. Reversible Contrast in Focus Series of Annular Bright Field Images of a Crystalline LiMn2O4 Nanowire. *Ultramicroscopy* **2013**, *125*, 43–48. https://doi.org/10.1016/j.ultramic.2012.09.011.

(26) Kimoto, K.; Asaka, T.; Yu, X.; Nagai, T.; Matsui, Y.; Ishizuka, K. Ultramicroscopy Local Crystal Structure Analysis with Several Picometer Precision Using Scanning Transmission Electron Microscopy. *Ultramicroscopy* **2010**, *110* (7), 778–782. https://doi.org/10.1016/j.ultramic.2009.11.014.

(27) Sato, Y.; Miyauchi, R.; Aoki, M.; Fujinaka, S.; Teranishi, R.; Kaneko, K. Large Electric-Field-Induced Strain Close to the Surface in Barium Titanate Studied by Atomic-Scale In Situ Electron Microscopy. *Phys. status solidi – Rapid Res. Lett.* **2020**, *14* (1), 1900488. https://doi.org/10.1002/pssr.201900488.

(28) Fujinaka, S.; Sato, Y.; Teranishi, R.; Kaneko, K. Understanding of Scanning-System Distortions of Atomic-Scale Scanning Transmission Electron Microscopy Images for Accurate Lattice Parameter Measurements. *J. Mater. Sci.* **2020**, *55* (19), 8123–8133. https://doi.org/10.1007/s10853-020-04602-w.

(29) De Backer, A.; van den Bos, K. H. W.; Van den Broek, W.; Sijbers, J.; Van Aert, S. StatSTEM: An Efficient Approach for Accurate and Precise Model-Based Quantification of Atomic Resolution Electron Microscopy Images. *Ultramicroscopy* **2016**, *171*, 104–116. https://doi.org/10.1016/j.ultramic.2016.08.018.

(30) van Dyck, D. High-Resolution Electron Microscopy; 2002; pp 105–171. https://doi.org/10.1016/S1076-5670(02)80062-3.

(31) *Springer Handbook of Condensed Matter and Materials Data*; Martienssen, W., Warlimont, H., Eds.; Springer Berlin Heidelberg, 2005; Vol. 66. https://doi.org/10.1007/3-540-30437-1.

(32) Aso, K.; Shigematsu, K.; Yamamoto, T.; Matsumura, S. Detection of Picometer-Order Atomic Displacements in Drift-Compensated HAADF-STEM Images of Gold Nanorods. *Microscopy* **2016**, *65* (5), 391–399. https://doi.org/10.1093/jmicro/dfw018.

(33) Rasmussen, C. E.; Williams, C. K. I. *Gaussian Processes for Machine Learning.*; 2004; Vol. 14. https://doi.org/10.1142/S0129065704001899.

(34) Hines, T. T.; Hetland, E. A. Revealing Transient Strain in Geodetic Data with Gaussian Process Regression. *Geophys. J. Int.* **2017**, *212* (3), 2116–2130. https://doi.org/10.1093/gji/ggx525.

(35) Young, T. III. An Essay on the Cohesion of Fluids. *Philos. Trans. R. Soc. London* **1805**, *95* (1), 65–87. https://doi.org/10.1098/rstl.1805.0005.

(36) Laplace, P. S.; Gordon, H.; Gordon, J. *Traité de Mécanique Céleste*; De L'Imprimerie de Crapelet : Chez J.B.M. Duprat: A Paris, 1798. https://doi.org/10.5479/sil.338664.39088005644752.

(37) Huang, Z. The Young-Laplace Equation Associated with Transverse Shear Stress within the Surface Layer of a Solid. **2018**, *0839*. https://doi.org/10.1080/09500839.2018.1528014.

(38) Mays, C. W.; Vermaak, J. S.; Kuhlmann-Wilsdorf, D. On Surface Stress and Surface Tension. *Surf. Sci.* **1968**, *12* (2), 134–140. https://doi.org/10.1016/0039-6028(68)90119-2.

(39) Landau, L. D.; Lifshitz, E. M. Fluid Mechanics: Landau and Lifshitz: Course of Theoretical Physics. *Image Rochester NY*. 1987, p 539. https://doi.org/10.1007/b138775.

(40) Wasserman, H. J.; Vermaak, J. S. On the Determination of a Lattice Contraction in Very Small Silver Particles. *Surf. Sci.* **1970**, *22* (1), 164–172. https://doi.org/10.1016/0039-6028(70)90031-2.

(41) Borisevich, A. Y.; Lupini, A. R.; Pennycook, S. J. Depth Sectioning with the Aberration-Corrected Scanning Transmission Electron Microscope. *Proc. Natl. Acad. Sci.* **2006**, *103* (9), 3044–3048. https://doi.org/10.1073/pnas.0507105103.

(42) Wales, D. J.; Doye, J. P. K. Global Optimization by Basin-Hopping and the Lowest Energy Structures of Lennard-Jones Clusters Containing up to 110 Atoms. *J. Phys. Chem. A* **1997**, *101* (28), 5111–5116. https://doi.org/10.1021/jp970984n.

**For Table of Contents Only**

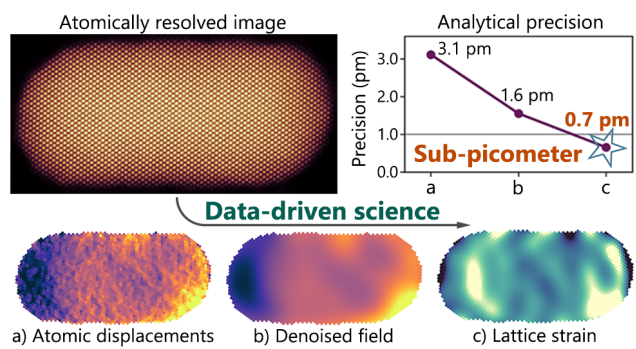

3.25 × 1.75 in., 300 dpi